%% file: main.tex
\documentclass[sigconf,nonacm]{acmart}
\pdfoutput =1
\usepackage{algorithmic}
\usepackage{graphicx}
\usepackage{textcomp}
\usepackage{xcolor}
\usepackage{colortbl}
\usepackage{tikz}

\usepackage{algorithmic}
\usepackage{graphicx}
\usepackage{textcomp}
\usepackage{xcolor}
\usepackage{longtable}
\usepackage{float}
\usepackage{graphicx}
\usepackage{multirow}
\usepackage{balance}
\usepackage{flushend}
\usepackage{hyperref}
\usepackage{tabularx} 
\usepackage{booktabs} 
\usepackage{hhline}
\usepackage{enumitem}
\usepackage{array}
\usepackage{lipsum}
\usepackage{multicol}
\usepackage{balance}
\AtBeginDocument{%
  \providecommand\BibTeX{{%
    \normalfont B\kern-0.5em{\scshape i\kern-0.25em b}\kern-0.8em\TeX}}}

\copyrightyear{2024}
\acmYear{2024}

\acmConference[MSR'2024: Registered Reports]{Mining Software Repositories (Registered Reports)}{April 15--16,
  2024}{Lisbon, Portugal}
\acmBooktitle{MSR '2024: Mining Software Repositories (Registered Reports)} 

\begin{document}

\newenvironment{boxedtext}
    {
     \begin{center}
    \renewcommand{\arraystretch}{1.2}
    \begin{tabular}{|p{0.96\linewidth}|}
    \hline
    }
    { 
    \\ \hline
    \end{tabular} 
    \renewcommand{\arraystretch}{1}
    \end{center}
       }

\newcommand{\revision}[1]{{#1}}
\newcommand{\code}[1]{ { \tt #1}}

\title[Influence of Toxic and Gender Discriminatory Communication on Perceptible Diversity]{Assessing the Influence of Toxic and Gender Discriminatory Communication on Perceptible Diversity in OSS Projects}


\author{Sayma Sultana$^\heartsuit$, Gias Uddin$^\clubsuit$, Amiangshu Bosu$^\heartsuit$}
\affiliation{%
  \institution{$^\heartsuit$Wayne State University, Detroit, Michigan, USA}
  \institution{$^\clubsuit$York University, Toronto, ON, Canada}
  \country{}
\city{\textit{sayma@wayne.edu, guddin@yorku.ca, amiangshu.bosu@wayne.edu}}}



\renewcommand{\shortauthors}{Sultana et al.}

\begin{abstract}
\input{Sections/abstract}
\end{abstract}




\maketitle
\balance
\input{Sections/introduction}

\input{Sections/background}
\input{Sections/hypotheses}
\input{Sections/execution_plan}


\bibliographystyle{ACM-Reference-Format}

\bibliography{references}


\appendix

\end{document}

%% file: Sections/abstract.tex
The presence of toxic and gender-identity derogatory language in open-source software (OSS) communities has recently become a focal point for researchers. Such comments not only lead to frustration and disengagement among developers but may also influence their leave from the OSS projects. Despite ample evidence suggesting that diverse teams enhance productivity, the existence of toxic or gender identity discriminatory communications poses a significant threat to the participation of individuals from marginalized groups and, as such, may act as a barrier to fostering diversity and inclusion in OSS projects. However, there is a notable lack of research dedicated to exploring the association between gender-based toxic and derogatory language with a perceptible diversity of open-source software teams. Consequently, this study aims to investigate how such content influences the gender, ethnicity, and tenure diversity of open-source software development teams. To achieve this, we extract data from active GitHub projects, assess various project characteristics, and identify instances of toxic and gender-discriminatory language within issue/pull request comments. Using these attributes, we construct regression models to explore how they associate with the perceptible diversity of those projects.

%% file: Sections/introduction.tex
\section{Introduction}
\label{sec:intro}
 According to the 2022 Stack Overflow 
survey, more than 90\% of professional software developers worldwide identify as men~\cite{stackoverflow_survey_2022}. 
As a result, the demographic of software users is vastly different from those developing the software. Lack of diversity among contemporary computing organizations not only lowers software development productivity~\cite{vasilescu2015gender,foundjem-icse-2021} but also creates software that is biased against various minorities~\cite{padala-2022,ledford2019millions}. Prior research shows that
diverse teams can better understand and address the needs of a diverse user base, as members from diverse backgrounds bring a range of perspectives, experiences, and ideas that can lead to more innovative and creative solutions to complex problems~\cite{mannix2005differences}. As software
touches all aspects of modern life, from communication to entertainment to transportation to healthcare
to education to finance to home appliances providing convenience, it is crucial that software serves the needs of all groups of users and reflects the diversity of our society. Unfortunately, most software projects severely lack diversity~\cite{bosu2019diversity,sultana2023code}.

Diversity issues in Open Source Software (OSS) projects have been subjected to numerous prior studies. Although OSS communities are increasingly becoming more diverse due to various Diversity, Equity, and Inclusion (DEI) initiatives~\cite{mozilla-diversity,fedora-diversity}, a recent study reports the ratios of women no more than 9\% among popular OSS projects ~\cite{sultana2023code}. Thomas \textit{et} al. showed that African American women have encountered complete isolation in the domain of computer science, and they do not know if this discrimination occurs due to their race or gender~\cite{thomas2018speaking}. Studies also reported that in useR! Conference for R users, there is an underrepresentation of non-white attendees~\cite{bollmann2017first}. 

Although OSS projects, as well as commercial organizations, have taken DEI initiatives to build diverse teams~\cite{mozilla-diversity,fedora-diversity,dagan2023building}, various implicit and explicit biases hinder onboarding and participation of minorities~\cite{sultana2023code,imtiaz2019investigating,wang2019implicit}. Not only biases, but the results of recent studies also found negative interactions among developers as a significant barrier against DEI~\cite{gunawardena2022destructive,murphy2022pushback}, since anti-social communication, such as pushback and destructive criticisms, disproportionately hurt minorities~\cite{gunawardena2022destructive,murphy2022pushback}. 
While recent studies have characterized various other anti-social communication among software developers using various lenses such as toxicity~\cite{raman2020stress,miller2022did,sarker2023automated}, incivility~\cite{ferreira2022incivility}, and gender identity-based derogation~\cite{sultana2023code}, their influence on DEI remains unexplored. An empirical investigation to assess the influence of toxic and gender identity derogatory communication on diversity is a crucial need since a blog written by an anonymous developer 
after leaving a toxic OSS community mentions suggest severe negative consequences on DEI, \textit{“..it’s time to do a deep dive into the mailing list archives or
chat logs. ... Searching for terms that degrade women (chick, babe, girl, bitch, cunt), homophobic slurs used as negative feedback (“that’s so gay”), and ableist terms (dumb, retarded, lame), may allow you to
get a sense of how aware (or not aware) the community is about the impact of their language choice on minorities.”}~\cite{anonymous_FOSS}.

However, prior studies on toxicity and gender discriminatory language have primarily focused on two fronts: the automated identification of unhealthy interaction~\cite{raman2020stress}, toxicity~\cite{sarker2023automated}, incivility~\cite{ferreira2022incivility}, sexist and misogynistic language~\cite{anzovino2018automatic, sultana2023code} and the exploration of toxicity across various platforms such as Wikipedia, GitHub, StackOverflow, Twitter (X), YouTube, and others. Limited attention has been given to understanding the long-term consequences of such interactions in building a diverse OSS community.  Maintainers of OSS projects note that facing negativity can have both mental and operational repercussions, potentially leading to project abandonment or stepping down from leadership roles.  Destefanis et al. \cite{destefanis2016software} discovered that impolite issue comments correlate with slower resolutions, and disengagement is a consequence of the lack of peer support and overall dissatisfaction \cite{graziotin2018happens, miller2019people}. Additionally, attracting and retaining contributors~\cite{crowston2008free}, especially from underrepresented groups~\cite{bosu2019diversity}, has been identified as a challenge in the realm of OSS projects.  

Despite repeated reports of the negative influence of toxicity and gender identity-based derogations~\cite{singh2022discrimination}, we are missing empirical evidence regarding how the prevalence of toxicity and gender-derogatory communication associate the diversity of various OSS communities. While such an investigation may turn out to be confirmatory, it may provide empirical evidence and quantification to support and motivate more OSS projects to adopt initiatives to promote inclusive communication. Additionally, a time-series analysis of toxicity and gender-derogatory communication against diversity measures for a project may provide further evidence regarding changes in diversity with toxicity and gender identity-based derogations. 
Hence, this study aims to fill this research gap by assessing \textit{the association between the toxic and gender-derogatory content's prevalence and perceptible diversity among OSS projects.}

While diversity has many facets, such as age, tenure, gender identity, race, culture, ethnicity, religion, political belief, and sexual orientation, most demographic factors cannot be reliably determined without input from a person. However, recent studies have used automated tools to perceive gender~\cite{sultana2023code,bosu2019diversity,vasilescu2015gender,imtiaz2019investigating}, project tenure~\cite{vasilescu2015gender}, and ethnicity~\cite{weeraddana2023empirical} from a persons name and avatar. Therefore, our proposed study focuses on these three perceptible demographics. 

 For this purpose, we will collect a dataset of OSS projects from GitHub that satisfies our sampling criteria. We will mine all issues and pull request comments for those projects for the 2023 calendar year. Using two state-of-the-art SE domain-specific   toxicity~\cite{sarker2023automated} and gender derogatory~\cite{sultana2023code} detection tools, we will automatically identify toxic and gender discriminatory texts. We will train multivariate regression models, where three diversity indices measured using the Blau/Simpson index would be dependents, and the ratio of toxic /gender discriminatory texts would be one of the independents. To account for confounding factors, we will use various project characteristics such as the age of a project, the number of contributors, the number of total commits, the number of issues, the number of releases, the number of pull requests, and if the project has a code of conduct, as independents. 
 We will assess the performance of the models using Adjusted $R^2$ and log-likelihood tests. We will assess the association between perceptible diversity and the prevalence of toxicity /gender-derogatory texts using regression coefficients and their significance (i.e., p-value <0.05).
Primary expected contributions from this proposed study are as follows:
 \begin{itemize}
     \item { Empirical evidence regarding associations between the prevalence of toxic and gender discriminatory communication and perceptible gender, tenure, and ethnic diversity of OSS projects hosted on GitHub.}
     \item {Longitudinal analyses of variations in perceptible diversity against changes in toxic and gender discriminatory communication.}
 \end{itemize}

%% file: Sections/background.tex
\section{Backgroud}
The following subsections briefly describe the key terminologies associated with this proposed study.

\subsection{Toxicity}
While both researchers from NLP and SE have studied the toxicity phenomenon, the definition of toxicity differs and is often customized to fit the specific context. 
Miller et al. claim that a variety of antisocial activities, including hate speech, trolling, flaming, and cyberbullying, fall under the category of Toxicity \cite{miller2022did}. Sarker et al.'s definition is more expansive to fit multicultural workplace communication such as OSS projects, and it includes offensive name-calling,  insults, threats, personal attacks, flirtations,  reference to sexual activities, and swearing or cursing~\cite{sarker2023automated}. This study follows Sarker et al.'s definition since their labeled dataset is the largest one for the SE domain, and their tool named ToxiCR~\cite{sarker2023automated} is the current SOTA for toxicity detection from SE texts.

\subsection{Gender identity-based discrimination}

Prior works from the NLP domain were primarily discriminatory sexism and misogyny, and such tools often fail to identify prejudice or discrimination towards LGBTQ+ persons. In their work, Sultana et al. developed SGID4SE to identify any text expressing prejudice or discrimination based on a person's gender, biological sex, gender identity, or sexual orientation.
While flirtations and gender-based insults and identity attack falls into the definition of sexism and derogatory content based on gender, several other types of discriminatory content have been found in the communication channels of OSS projects that are directed towards women and LGBTQ+ people~\cite{squire2015floss, sultana2023code}. Such types of gender-discriminatory content pose barriers against the participation of women and LGBTQ+ people and negatively hampers DEI initiatives.

\subsection{Measuring diversity}
Researchers from biology have established several quantitative measures, a.k .a. the diversity index, to measure biodiversity.  \textit{Richness} is the most straightforward measure that quantifies the number of distinct categories, with 1 indicating all members belonging to the same category~\cite{dejong1975comparison}.
\textit{Evenness} describes how close in numbers each species in an environment is.~\cite{dejong1975comparison}.
 Blau index aka Simpson's index~\cite{simpson1949measurement}, measures the probability that two randomly selected subjects will be from different categories. In the absence of diversity (1 species), the probability that two individuals randomly selected belonging to different groups would be 0. 

%% file: Sections/hypotheses.tex
\section{Hypotheses}
\label{sec:hypotheses}

We aim to investigate whether the prevalence of toxic or discriminatory content influences various perceptible diversities of an OSS project.  Therefore, our first research question is:

 \begin{boxedtext}
\textbf{RQ1:} \emph{
 Does the ratio of toxic texts among issue/pull request comments influence the perceptible diversity of open-source software projects?}
\end{boxedtext}

While researching toxicity in open-source software projects, Sarker et al. \cite{sarker2023automated} provide a rubric for identifying toxic comments for the domain of software developers. They mention that ``Attacking a person’s identity
(e.g., race, religion, nationality, gender, or sexual orientation) would be marked as ‘toxic’''. Therefore, toxic comments can influence the participation of developers from different races, nations, genders, and sexual orientations. Researchers have conducted studies to investigate gender and ethnic diversity in teams of software developers~\cite{weeraddana2023empirical}. Vasilescu et al. ~\cite{vasilescu2015gender} studied the relationship of gender and tenure diversity with team productivity and turnover rate. \revision{Here, we take the ratio of toxic comments to total comments of a particular project and investigate the influence of toxic content on gender, ethnicity, and tenure diversity. }
We formalize the first research question into the following three hypotheses, with each one having a null and alternate. 

\boldmath{$H1.1_0$:} \textit{There is no significant association between the gender diversity index and the ratio of toxic texts in an OSS project. }

\boldmath{$H1.1_a$:} \textit{There is a significant association between the gender diversity index and the ratio of toxic texts in an OSS project.}

\boldmath{$H1.2_0$:} \textit{There is no significant association between the ethnic diversity index and the ratio of toxic texts in an OSS project. }

\boldmath{$H1.2_a$:} \textit{There is a significant association between the ethnic diversity index and the ratio of toxic texts in an OSS project.}

\boldmath{$H1.3_0$:} \textit{There is no significant association between the tenure diversity index and the ratio of toxic texts in an OSS project. }

\boldmath{$H1.3_a$:} \textit{There is a significant association between the tenure diversity index and the ratio of toxic texts in an OSS project.}

Prior studies also found sexist, misogynistic, and gender-discriminatory content in the communication excerpts of open-source software communities~\cite{singh2022discrimination, sultana2023code}. Therefore, we want to investigate if gender or sexual orientation-based discriminatory content influences gender, ethnic, and tenure diversity and formulate our second research question as:
 \begin{boxedtext}
\textbf{RQ2:} \emph{
 Does the ratio of gender identity discriminatory texts among issue/pull request comments influence the perceptible diversity of open-source software projects?}
\end{boxedtext}

Since such type of discriminatory content specifically targets women or people from different sexual orientations, it also might target women of different races. We derive the following hypotheses from the research question above. 

\boldmath{$H2.1_0$:} \textit{There is no significant association between the gender diversity index and the ratio of gender discriminatory texts in an OSS project.}

\boldmath{$H2.1_a$:} \textit{There is a significant association between the gender diversity index and the ratio of gender discriminatory texts in an OSS project.}

\boldmath{$H2.2_0$:} \textit{There is no significant association between the ethnic diversity index and the ratio of gender discriminatory texts in an OSS project. }

\boldmath{$H2.2_a$:} \textit{There is a significant association between the ethnic diversity index and the ratio of gender discriminatory texts in an OSS project.}

\boldmath{$H2.3_0$:} \textit{There is no significant association between the tenure diversity index and the ratio of gender discriminatory texts in an OSS project. }

\boldmath{$H2.3_a$:} \textit{There is a significant association between the tenure diversity index and the ratio of gender discriminatory texts in an OSS project.}

Due to increased levels of toxicity or gender discriminatory texts, some people may leave a project for good~\cite{anonymous_FOSS}. Since minorities are more likely to be marginalized and become victims, an exodus of minorities will lower diversity. Our final research question aims to analyze such association with a time-series-based analysis. 

  \begin{boxedtext}
 \textbf{RQ3:} \emph{
  Do increased ratios of toxicity /gender discriminatory texts associate with decreased perceptible diversity within certain time intervals and vice versa? }
 \end{boxedtext}

 The three alternate hypotheses are.

 \boldmath{$H3.1_a$:} \textit{There is a significant association between the time series representing ratios of toxic /gender discriminatory texts and the series representing gender diversity index with some lag.}

 \boldmath{$H3.2_a$:} \textit{There is a significant association between the time series representing ratios of toxic /gender discriminatory texts and those representing ethnic diversity index with some lag.}

 \boldmath{$H3.3_a$:} \textit{There is a significant association between the time series representing ratios of toxic /gender discriminatory texts and the series representing tenure diversity index with some lag.}

%% file: Sections/execution_plan.tex
\section{Execution Plan}
\label{sec:execution-plan}
We will test our hypotheses by designing multinomial regression models. Table \ref{table:variable_descriptions} summarizes the dependent and independent variables we will utilize for our regression modeling. We discuss the variables below.
\subsection{Variables}
\input{Sections/variables}

\subsection{Dataset}
We intend to conduct our study on open-source projects hosted on GitHub. We employed the GitHub search tool developed by Dabic \textit{et} al.~\cite{dabic2021sampling} for project selection. This tool allows for selecting projects based on various criteria, such as the number of contributors, forks, commits, stars, programming language, and more. Following the recommendations of Kalliamvakou \textit{et} al.~\cite{kalliamvakou2016depth}, we will filter projects based on four criteria. The first two criteria enable the selection of software projects with active communication among contributors, ensuring sufficient data for identifying toxic and derogatory content.  The remaining criteria assist in narrowing down the search space:

\begin{itemize}
    \item Uses one of the primary languages: Java, C, C++, Python, JavaScript, C\#, Go, PHP, Typescript,  and Ruby. {These languages are the top ten programming languages on GitHub.}
    \item Has at least 20 contributors and is publicly available with an open-source license. 
    \item Has at least 20 PRs. 
    \item Has at least ten stars. 
\end{itemize}

We will also consider only non-forked projects since forked projects have a similar project history as their main ones and thus can bias our result. We also plan to divide our dataset into four groups based on the number of developers to have an in-depth idea about the occurrence and influence of toxic and derogatory language following prior study on gender bias~\cite{sultana2023code}. We estimate approximately 50,000 projects to survive this step.

\subsection{Datatset preparation}
\label{dataset-prepare}
We will write Python Scripts using the PyGitHub library~\cite{pygithub} to download and count the number of pull/ issue requests for the filtered projects. For RQ1 and RQ2, we will only focus on the data from the 2023 calendar year. We make this decision for two reasons. First, since diversity is a snapshot in time, a project's snapshot from five years earlier may not match the current one. Second, due to changes in leadership and governance, the prevalence of toxicity may change over the years. 
While prior studies have used a snapshot for six months, we take one year as the interval since toxic interactions, although they have significant consequences, are rare (less than 1\%). Therefore, a smaller sample may not provide adequate data points for testing for smaller projects.

After data mining, we will further exclude projects that did not have at least 100 pull requests and 20 different contributors in 2023.

\subsubsection{Mining \& cleaning dataset} 
 We plan to investigate the gender and ethnic diversity of the projects. Therefore, we will remove all contributions to the projects by non-human users. Prior studies have used GitHub API to filter out the contribution of bots~\cite{weeraddana2023empirical}. If a human contributes, the API returns the type as USER. So, we will filter out all other types.

\subsubsection{Identifying toxic \& derogatory content} We will use ToxiCR~\cite{sarker2023automated} and SGID4SE~\cite{sultana2023code} for automated identfications. Following the recommendation of prior study~\cite{novielli2018benchmark}, we plan to validate the identified toxic comments from communication excerpts. {For that purpose, we will randomly select $N$ toxic comments identified by the tool and annotate them independently by at least two researchers. The value of $N$ will be computed based on Taro Yamane's formula~\cite{yamane1973statistics} to satisfy a 95\% confidence interval and 5\% error margin. 
Similarly, we will also conduct a validation step for the identified derogatory content. We will randomly select $N$ comments from the positive class and annotate manually to assess the reliability of the tools (i.e., whether the tools' performance degraded and their classification can be used for our study). } 
{Since ToxiCR and SGID4SE are trained for texts written in English, we will use the langdetect~\cite{langdetect} Python library to filter out non-English texts.}

\subsubsection{Calculating variables} Our study also requires a few other project attributes. We will calculate those attributes using Python scripts.

\subsection{Analysis Plan} We will use \textit{multivariate regression modeling} techniques for our model development. Linear and logistic regression are widely utilized methods to explore the relationship between a dependent variable and one or more independent variables, especially when the dependent variable is scalar or binary, respectively~\cite{Regression}. Since the value for `Blau index' is scalar, we will use linear regression models to answer the first two research questions. Table \ref{table:variable_descriptions} lists all the dependent and independent variables necessary for our models.

Recent studies in Software Engineering~\cite{mcintosh,bosu_2, sultana2023code} have motivated us to adopt Harrell Jr.'s approach for constructing and analyzing models to validate our proposed hypotheses~\cite{harrell2015regression}. By following Harrell's methodology, we can effectively capture non-linear relationships among variables while addressing concerns about overfitting, where the model may perform exceptionally well on the training dataset but poorly on other data~\cite{harrell2015regression}. To implement Harrell's regression techniques, we will utilize the \textit{rms} package in R~\cite{R_harrell}. A brief explanation of our model development and analysis approach follows.

\subsubsection{Correlation \& redundancy analysis: } We will perform correlation analysis on the independent variables within the models to eliminate any correlated factors. For that purpose, we will use the Spearman rank correlation test (\(\rho\)), known for its resilience to datasets that deviate from a normal distribution. Additionally, we will identify and exclude redundant variables by creating a hierarchical overview of correlated variables. Within the same sub-hierarchy, variables with an absolute Spearman correlation coefficient (\(\rho\)) greater than 0.7 will be assessed, and one will be selected for inclusion in the final regression model. Previous studies in software engineering have also adopted 0.7 as the threshold value for identifying redundant variables~\cite{mcintosh}~\cite{Thongtanunam}~\cite{bosu_2}.

\subsubsection{Normality adjustment: } We will assess the normal distribution of our response or dependent variable. If they deviate from normality, we will apply a log transformation, denoted as \textit{ln(x)}, to these variables, as suggested by prior research~\cite{mcintosh}.

\subsubsection{Equation for regression model: } {To illustrate, the linear regression for testing H1.1 is as follows.}

\begin{equation}
  \begin{aligned}[b]
  & Blau_{Gender} \thicksim projectAge + isCorporateBacked  \\
  & +  ln(numOfContributor) + isGaming \\
  & + ratioOfToxicComment + hasCoC + numberIssues\\
  & + ln(numOfCommits) + numberOfBuilds +
  numberPRs 
  \end{aligned}
\end{equation}

{We can obtain equations for $H1.2$  and $H1.3$ by replacing $Blau_{Gender}$ with $Blau_{Tenure}$ and $Blau_{Ethnicity}$ in equation 1, respectively.
 On the other hand, by replacing the independent variable \\ \code{ratioOfToxicComment} with \code{ratioOfDicriminatoryComment} in equation 1, we obtain the equation for testing H2.1, as follows.}

\begin{equation}
  \begin{aligned}[b]
  & Blau_{Gender} \thicksim projectAge + isCorporateBacked  \\
  & +  ln(numOfContributor) + isGaming \\
  & + ratioOfDicriminatoryComment + hasCoC + numberIssues\\
  & + ln(numOfCommits) + numberOfBuilds +
  numberPRs 
  \end{aligned}
\end{equation}

{Similarly, We can obtain equations for $H2.2$  and $H2.3$ by replacing $Blau_{Gender}$ with $Blau_{Tenure}$ and $Blau_{Ethnicity}$ in equation 2, respectively.}

\subsubsection{Assessment for model performance: } We will compute McFadden's \(R^2\) to evaluate the goodness of fit of our models to the datasets. Additionally, we will check if the model has significant explanatory power over a NULL model using the log-likelihood test~\cite{harrell2015regression}.

\subsubsection{Estimate the power of independent variables of interest: } {We will utilize the p-value to assess the significant association of independents with the dependent under consideration, with values less than 0.05 indicating significant ones. We will use the regression coefficients to estimate how an independent may change the dependent. Finally, we will estimate the explanatory power of an independent, as variance explained, using the approach proposed by Chambers and Hastie~\cite{chambers2017statistical} and implemented in the \code{anova} method of the \code{stats} package in R. We will use Cohen's $f^2$ to estimate the effect sizes, with $f^2 \ge   0.02$, $f^2  \ge 0.15$, and $f^2 \ge  0.35$ representing small, medium, and large effect sizes, respectively~\cite{selya2012practical}. }

\subsubsection{Examination of the independent variables in the outcome: } For the linear regression models, we will create plots illustrating the variation of dependent variables against the change in an independent variable at a time while holding other independent variables constant at their median values. This graphical representation can provide insights into how alterations in any individual independent variable, such as the ratio of toxicity or gender identity discriminatory texts, influence a particular diversity index.

\subsubsection{Acceptance/rejection of hypotheses:}{After training a linear regression model, . If our results suggest a reliable model, we will check whether the independent variable of interest has a significant coefficient power in this model (i.e., $p <0.05$). For example, for $H1.1$, we will use   $Blau_{Gender}$ as the dependent variable, and \code{ratioOfToxicComment} is the independent variable we aim to examine. We accept the alternate hypothesis if we obtain a fitted model (i.e., significantly better than a null model), and \code{ratioOfToxicComment} has a significant association.  If not significant, we accept the null hypothesis. Similarly, for $H1.2$\\ \code{ratioOfDiscriminatoryComment} would be the independent variable of interest }.

\subsection{Time series analysis (RQ3)}
{Each time series analysis is specific to a particular project and should not be done in aggregation. The result for each project may differ in terms of association as well when such association becomes most prominent.  However, a project must have a minimum number of data points to identify any significant association, which experts on time-series analysis suggest to be at least 40-50 \cite{poole2002hypothesis,warner1998spectral}.  Therefore, from the projects surviving filtering steps described in Section~\ref{dataset-prepare}, we will select only the ones that have been active for at least ten years (i.e., 40 quarters/data points ). If the number of projects surviving this step exceeds 400 (i.e., a number required to satisfy a 95\%  confidence interval and 5\% error margin), we will randomly select 400.
We will download the entire issue/pull history for these selected projects. We will compute diversity indices and ratios of toxicity /gender discriminatory texts for each quarter. } Then, we will compute correlations between these two time series. However, this analysis will require some extra steps.

Autocorrelation is the degree of similarity between a given time series and a lagged version of itself over successive time intervals. Without correcting autocorrelations, we may observe spurious relationships~\cite{yule1926we}. We will use Augmented Dickey-Fuller~\cite{said1984testing} tests using the \emph{tseries}-R package to check if those series are auto-correlated. If yes, we will follow the suggestions of Farnum and Stanton~\cite{farnum1989quantitative}  and introduce ``first differences'' ~\cite{brockwell2013time} for each time series. For example, if $S_m$ indicates the number of posts in $m^{th}$ quarter, then $\Delta S_m= S_m -S_{m-1} $.
Using the {\textit{ccf}} function from the \emph{stats}-R package, we compute cross-correlations, i.e., the measure of similarity of two-time series as a function of a time-lag applied to one of them. 

\subsection{Validity Threats} 

\textbf{Internal Validity} We intend to investigate our hypotheses across a sample of GitHub projects, recognizing that the project selection process could introduce internal validity concerns to our study. In alignment with the recommendation by Kalliamvakou \textit{et} al.~\cite{kalliamvakou2016depth}, we have specifically chosen projects that meet certain criteria: usage of one of the programming languages, a minimum of 20 contributors, and 20 pull requests (PRs), and a minimum of 10 stars. These criteria aim to exclude discontinued or small projects that may not offer valuable insights. Additionally, we intend to implement a stratified sampling strategy in project selection to analyze the presence and impact of toxic and derogatory language across different project sizes. Despite our meticulous study design, it is important to acknowledge that the characteristics of our sample may not perfectly represent the entire GitHub ecosystem.

\textbf{External Validity} Our investigation is confined to open-source GitHub projects, and the conclusions drawn may not be generalizable to projects in other OSS domains. {To promote replications of this study on various other contexts, We will make our scripts and deidentified datasets publicly available on Zenodo with a permanent DOI.}

\textbf{Construct Validity} The major construct that poses a threat to this study is resolving the gender and ethnicity of the developers using the existing gender and ethnicity detection tools. {We limit analyzing the influence of gender diversity within men and women since other types of gender and sexual orientation can not be identified from the names.} Moreover, gender and ethnicity detection tools are not beyond limitations. GenderComputer classifies takes names and locations as input and classifies those into four groups: (i) male, (ii) female, (iii) unisex, (iv) unknown. We will consider both unisex and unknown as not-resolved since we limit our study to only men and women. Santamaría and Mihaljević evaluated the performance of the Gender-Guesser tool and found that 20.12\% names can not be resolved by this tool~\cite{santamaria2018comparison}. Therefore, being motivated by prior study~\cite{weeraddana2023empirical}, we tried to limit the number of unresolved genders using the combination of three gender identifier tools. 

Also, identifying the toxic and gender-discriminatory content using the existing tools introduces threats. ToxiCR reported having 95.8\% accuracy in their dataset, and the gender discriminatory content identification tool has 95.7\% accuracy with 85.9\% precision. We will conduct validation steps for both toxic and gender-discriminatory content to reduce the number of misclassifications of positive classes.

\textbf{Conclusion validity} We will employ regression-based models for analysis using established and mature libraries such as \code{stats} and \code{rms} for model training. Standard metrics, such as the coefficient ($Z$) for linear models, will be employed to estimate effects. As a result, it is improbable that any threats to validity will emerge from the evaluation metrics, library choices, or assessment of dependent variables.

%% file: Sections/variables.tex
In this section, we present the detailed process of how those variables will be calculated. We choose different attributes of open-source projects that might influence the diversity factor and have been used by prior studies to study different types of diversity. We do not describe the computation of metrics that can be directly mined from GitHub (i.e., number of commits/ pull requests/ issues/ releases)

\begin{table*}
	\caption{Descriptions and rationale of the dependent and independent variables in our regression models}
	\centering \label{table:variable_descriptions}
	\input{Table/variables}
	\vspace{-8pt}
\end{table*}

\subsubsection{Project Age:} We will take the difference of months between the time of the last pull request merge and the time of project creation.

\subsubsection{Type of project sponsor:} For this attribute, we will consider the type of administrators who maintain the project: whether the project is maintained by a corporate company or not. A prior study showed that the corporate company-backed model shows less toxicity~\cite{raman2020stress}.


\subsubsection{IsGaming:} It refers to the domain of the project, whether that is gaming software or other types of software. Prior studies show that gaming software harbors more toxicity and derogatory content~\cite{miller2022did, sultana2023code}.

\subsubsection{Ratio of toxic comments :} The Perspective API, an advanced toxicity detector by Google, is widely recognized for evaluating online communications. However, previous studies have indicated that Natural Language Processing (NLP) tools designed for general domains exhibit subpar performance when applied to datasets within the Software Engineering domain. Research by Sarker et al. demonstrated that tools like STRUDEL (developed by Raman et al.~\cite{raman2020stress}) and several off-the-shelf toxicity detectors are not reliably effective for the Software Engineering dataset. In contrast, ToxiCR, developed by Sarker and colleagues, achieves impressive results with a 95.8\% accuracy and 88.9\% F1-score, surpassing the performance of other tools. Consequently, our approach involves utilizing ToxiCR\cite{sarker2023automated} to identify the prevalence of toxic comments in issue and pull request comments.

\subsubsection{Ratio of gender discriminatory comment:} Though there have been numerous studies to identify misogynistic content in the online communication medium, e.g., Twitter, YouTube comments, and so on, research to identify misogynistic text in the domain of software developers is at the outset. Sultana and her colleagues developed a tool to identify derogatory content based on gender and different sexual orientation~\cite{sultana2023code}. Their BERT-based model
achieved the best performance with 85.9\% precision and 82.9\% F1-Score for the positive class. with overall 95.7\%
accuracy. We plan to leverage that tool to identify the number of gender discriminatory content~\cite{sultana2023code}. 

\subsubsection{If there is a code of conduct:} Many projects adopt a code of conduct to maintain healthy interaction among the project contractors. So, we will check if projects have a formal code of conduct in their repository.




\subsubsection{Blau Index:} Prior studies have used Blau index/diversity index/Simpson's index~\cite{simpson1949measurement} to understand the diversity of open-source projects~\cite{weeraddana2023empirical, vasilescu2015gender}. We will also use this metric for measuring gender and ethnic diversity. The metric ranges from 0 to 1. The lower the value is for Blau, the less diverse the community is.  The formula for calculating the metric is shown below:

\[ Blau = 1- \sum_{i=1}^{S} (n_i/ N) ^2  \]

For example, for a team of 15 (N= 15) people, if there are six people from the perceptible Asian group, two people are perceptibly Black, two people are perceptible Hispanic, and the rest of the people are White, then the ethnic diversity index for this group is, 1 - {$(6/15)^2$ +  $(2/15)^2$ + $(2/15)^2$ + $(5/15)^2$} = 0.69; We will calculate the Blau index for gender, ethnic and tenure diversity of the projects. {We express the Blau Index for gender diversity with $Blau_{Gender}$ (equation 1). We will express ethnic and tenure diversity with $Blau_{Ethnic}$ and $Blau_{Tenure}$, respectively.}

\emph{Gender diversity ($Blau_{Gender}$):} We limit our study only to men and women since it is not possible to identify other types of gender from user names. Prior study~\cite{weeraddana2023empirical} have used GenderComputer\footnote{https://github.com/tue-mdse/genderComputer}, genderGuesser\footnote{https://pypi.org/project/gender-guesser/} and Wiki-GenderSort\footnote{https://github.com/nicolasberube/Wiki-Gendersort} to infer gender from people's names. {GenderComputer takes name and location as input. } We will also use a combination of these tools to infer the gender of the contributors. In addition to name-based resolution, we will use GitHub profile picture-based resolution adopted in our recent study~\cite{sultana2023code}.

\emph{Ethnic diversity ($Blau_{Ethnic}$):} {According to Fredrik Barth, ethnicity refers to a system of social grouping where others classify individuals as belonging to a particular category based on a shared cultural heritage~\cite{barth2010introduction}.} To resolve the ethnic identities of the developers, we will use Name-Prism that has been used by multiple prior studies\cite{nadri2021insights,nadri2021relationship, weeraddana2023empirical,else2022giant}. Name-Prism is a name-based classification tool that classifies people's names into the following categories: White, Black, Asian/Pacific Islander (API), Hispanic, American Indian/Alaskan Native (AIAN), and Mixed Race (2PRACE).

\emph{Tenure diversity ($Blau_{Tenure}$):} {Tenure refers to one's length of time in a particular space, e.g., project/GitHub \cite{vasilescu2015gender}.} To study the relationship of tenure with turnover and productivity, Bogdan et al. ~\cite{vasilescu2015gender} studied two types of tenure in OSS projects: (i) commit tenure, the global GitHub coding experience, and (ii) project tenure, the local project experience. Since we are studying the influence of toxic and gender-discriminatory content on a specific project, we focus on the project tenure of the contributors. {We will group developers' project tenure into the following categories: (i) Less than one year, (ii) 1 to 4 years, (iii) 5 to 9 years, and (iv) More than ten years.}


%% file: Table/variables.tex
\resizebox{\textwidth}{!}{%
\begin{tabular}{p{3.5cm} p {6cm} p {8cm}}
\hline
\rowcolor[HTML]{D9D9D9} 
\textbf{Name}           & \textbf{Description}                                & \textbf{Rationale}  \\ 

\rowcolor[HTML]{ffffff} 
\multicolumn{3}{c}{\cellcolor[HTML]{ffffff}\textbf{ Independent variables}}  \\   

\rowcolor[HTML]{EFEFEF} 
Project Age    &  Number of months project has started     &  Older projects showed higher level of toxicity~\cite{raman2020stress}\\

\rowcolor[HTML]{D9D9D9} 
If corporate backed   & If the project is backed up by corporate group or not  & Corporate projects show less toxicity than the non-corporate ones~\cite{raman2020stress} \\

\rowcolor[HTML]{EFEFEF} 
Number of contributors     &  The total number of contributors in a project in a period & Project with a large number of contributors can be more diverse and toxic ~\cite{miller2022did}  \\

\rowcolor[HTML]{D9D9D9} 
{IsGaming}  & If the project is from the gaming domain or  not &  Prior studies found that gaming projects use lots of profane keywords in issue discussions~\cite{miller2022did}\\

\rowcolor[HTML]{EFEFEF} 
Ratio of toxic comments &  Ratio of toxic comments and total comments for that project & Prior studies show related a phenomenon such as destructive criticisms hurt diversity~\cite{murphy2022pushback,gunawardena2022destructive}\\    

\rowcolor[HTML]{D9D9D9} 
Ratio of gender discriminatory comment  &  Ratio of gender discriminatory comments and total comments for that project during a period & A higher number of gender discriminatory content may adversely affect the participation of minorities and hurt diversity\\

\rowcolor[HTML]{EFEFEF} 
If there is a code of conduct  &  If the project has a code of conduct &  Establishment of code of conduct and its enforcement may discourage anti-social interactions and encourage minorities~\cite{singh2021codes}.     \\

\rowcolor[HTML]{D9D9D9} 
Number of commits     & The total number of commits in a period & Number of commits will show how much a project is active~\cite{weeraddana2023empirical}\\

\rowcolor[HTML]{EFEFEF}
Number of builds     & The total number of builds in a period &  Prior study used this attribute to study gender and ethnic diversity measuring how active a project is~\cite{weeraddana2023empirical}\\

\rowcolor[HTML]{D9D9D9} 
Number of issues &   The total number of issues for a period   &  Issue discussions may become heated due to unsatisfactory resolution~\cite{miller2022did}\\

\rowcolor[HTML]{EFEFEF}
Number of pull requests     & The total number of pull requests for a period &  Prior study used this attribute to study gender and ethnic diversity measuring how active a project is~\cite{weeraddana2023empirical}\\

\multicolumn{3}{c}{\cellcolor[HTML]{ffffff}\textbf{Dependent Variables}} \\

\rowcolor[HTML]{D9D9D9} 
Blau Index: Gender ($Blau_{Gender}$) & Gender diversity measure     &   Dependent variable to measure the association between toxicity/ gender discriminatory text and gender diversity ($H1.1$, $H2.1$ ) \\
\rowcolor[HTML]{EFEFEF}
Blau Index: Tenure ($Blau_{Tenure}$)  & Tenure diversity measure     &  Dependent variable to measure the association between toxicity/ gender discriminatory text and tenure diversity ($H1.2$, $H2.2$ )\\
\rowcolor[HTML]{D9D9D9} 
Blau Index: Ethnicity ($Blau_{Ethnicity}$)  & Ethnic diversity measure    &   Dependent variable to measure the association between toxicity/ gender discriminatory text and ethnicity diversity ($H1.3$, $H2.3$ ) \\

 \hline
\end{tabular}

}